\documentclass{llncs}
\usepackage{amssymb, graphicx}

\pdfoutput=1

\def\R{\Bbb R}

\begin{document}

\title{Euclidean Shortest Paths\\ in Simple Cube Curves at a Glance}
\author{Fajie Li and Reinhard Klette}
\institute{
Computer Science Department\\
The University of Auckland, New Zealand
}

\maketitle

\begin{abstract}
This paper reports about the development of two provably correct
approximate algorithms which calculate the Euclidean shortest path (ESP)
within a given cube-curve with arbitrary accuracy, defined by $\varepsilon >0$, and
in time complexity $\kappa(\varepsilon) \cdot {\cal O}(n)$, where
$\kappa(\varepsilon)$ is the length difference between the
path used for initialization and the minimum-length path, divided by
$\varepsilon$. A run-time diagram also illustrates this linear-time
behavior of the implemented ESP algorithm.
\end{abstract}

\section{Introduction}

Euclidean shortest path (ESP) problems are defined by
a (2D, 3D, ...) Euclidean space which contains (closed) 
polyhedral obstacles; the task is to compute a path which 
connects two given points in the space such that it
does not intersect the interior of any obstacle, and it
is of minimum Euclidean length. 

Examples are the ESP inside of a simple polygon, on the 
surface of a convex polytope,  or inside of a simply-connected 
polyhedron, or problems such as touring polygons, parts cutting, safari or 
zookeeper, or the watchman route. All-together, this defines a class
of immensely important computational problems of 
huge impact in economy, science or technology.

For time complexities of algorithms in this area, we cite two
examples. The general 3D ESP problem (e.g., path-planning in robotics)
is NP-hard, see J. Canny and J. H. Reif \cite{CAN1987}.

For 2D ESP problems, there are linear-time, but very complicated
algorithms (e.g., algorithms for ESP calculation in a simple polygon,
based on B. Chazelle's \cite{CHA1991}
triangulation of whole polygons), or  linear-time and easy-to-implement algorithms
(e.g., for the relative convex hull in the 2D grid, see
\cite{KLE1999}).

In this paper we consider ESPs in simple cube-curves, which are formed by
successively face-adjacent grid cubes (of the uniform orthogonal 3D grid,
see digital geometry \cite{KLE2004}).
T.\ B\"ulow and R.\ Klette published between 2000 and 2002 (see, e.g., 
\cite{BUE2002}) a so-called {\it rubberband algorithm} (RBA) 
for the  calculation of a Euclidean shortest path in a
simple cube-curve. \cite{BUE2002} stated two open problems: is this 
approximate RBA actually always converging (with numbers of iterations)
to the correct ESP, and is its time complexity actually linear as all experiments
indicated at that time.

This paper reports about the development of two 
appro\-xi\-mate RBAs, which always converge towards the ESP, and have
$\kappa(\varepsilon) \cdot {\cal O}(n)$  time complexity. The paper provides a first summary
of work done by F. Li and R. Klette in 2003 to 2006; for details and generalizations of 
the RBA approach for solving various ESP problems in 2D or 3D space, 
see \cite{LI2006,LI2007} and forthcoming publications of the authors.

\section{The Original RBA}

\begin{figure}[!h]
\begin{center}
\vspace{-3mm}
\includegraphics[width=57mm]{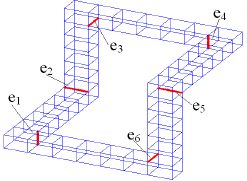}
\vspace{-3mm}
\end{center}
\caption{Critical edges $e_1$, $e_2$, $e_3$, $e_4$, $e_5$, and $e_6$.}
\label{fig:CrEd}
\end{figure}

{\it Critical edges} of a given cube-curve $g$ are those grid edges which are
incident with three cubes of the curve (see Figure~\ref{fig:CrEd}). Critical 
edges are the only possible locations for
vertices of an ESP \cite{KLE2000}. A subset of those will define the {\it step set}
of the RBA, which contains all those critical edges which contain exactly one 
ESP vertex each.

The {\it Original RBA}, as published in \cite{BUE2002,KLE2004}, is as follows:
it consists of two subprocesses, {\bf (i)} an initialization process (e.g., from an endpoint of
one critical edge to the closest endpoint of the subsequent
critical edge; satisfying a ``closed-path'' constraint at the end), and
{\bf (ii)} an iterative process which 
contracts the path during each of its loops, using a 
{\it break-off criterion}      
$${\cal L}_n - {\cal L}_{n+1} <  \varepsilon$$
where $\varepsilon > 0$, and 
${\cal L}_n$ is the total length of the path after the $n$th loop.

During each loop, the algorithm tries to shorten the path
locally by checking three options, called {\bf OP1}, {\bf OP2}, and {\bf OP3}.
{\bf OP1} and {\bf OP2} find the step set of critical edges. {\bf OP3} optimizes
the position of a vertex on its critical edge.
These options are defined as follows:

{\bf OP1}: delete vertex $p_i$  if  the line segment $p_{i-1}p{i+1}$ is in the 
           {\it tube} ${\bf g}$, which is the union of all 
           the grid cubes in the given simple cube-curve $g$;

{\bf OP2}: calculate intersection points of the triangle $p_{i-1}p_ip_{i+1}$
           with all critical edges (``between'' $p_{i-1}$ and  $p_{i+1}$)
           and replace  the subsequence $p_{i-1}, p_i, p_{i+1}$ by the 
           resulting convex arc, defined by these of intersection points;

{\bf OP3}: move $p_i$ on its critical edge $e$ into the optimum position 
          $p_{new}$,  with $d_e(p_{new},p_{i-1}) + d_e(p_{i+1},p_{new})
                         = \inf\{d_e(p, p_{i-1}) + d_e(p_{i+1},p):  p \in e\}$,
          where $d_e$ denotes the Euclidean distance.

We continue with vertices $p_{new},p_{i+1},p_{i+2}$ of the path. At the end
          of each loop we compare the total length of the new path with that of
          the path at the end of the previous loop.
          
\begin{figure}[!h]
\begin{center}
\vspace{-3mm}
\includegraphics[width=110mm]{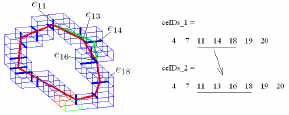}
\vspace{-3mm}
\end{center}
\caption{Illustration for the (original) Option 2.}
\label{fig:Op2}
\end{figure}

See Figure~\ref{fig:Op2} for {\bf OP2}. Here, vertices on critical edges $e_{11}$, $e_{14}$
and $e_{18}$ are replaced by a convex arc with vertices on critical edges $e_{11}$,
$e_{13}$, $e_{16}$, and $e_{18}$, and (in general) it may be $e_{11}$, $e_{14}$
and $e_{18}$ again within a subsequent loop  -- of course, for a
reduced length of the calculated path at this stage.

The situation with the original RBA in 2002 \cite{BUE2002} was as follows:
Even for very small values of  $\varepsilon$, the measured
time complexity indicated ${\cal O}(n)$, where $n$ is the number
of cubes in $g$. However, there was no
proof for the asymptotic time complexity of the original
RBA.
For a small number of test examples, calculated
paths seemed (!) to converge against the ESP. However,
no implemented algorithm for calculating the correct ESP was available, 
and (more general) no proof whether the path, provided 
by the original RBA, converges  towards the ESP.
Nevertheless, the algorithm is in use since 2002 (e.g.,  
in DNA research).

\section{Non-Existence of an Exact Arithmetic Algorithm}

An {\it arithmetic algorithm} consists of a finite number 
of steps of arithmetic operations, possibly also using 
input parameters from the field of rational numbers, 
using only the following basic operators:  $+ , -, \cdot ,  /$
or the $k$th root, for $k \ge 2$.

\begin{figure}[!t]
\begin{center}
\vspace{-3mm}
\includegraphics[width=60mm]{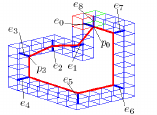}
\vspace{-3mm}
\end{center}
\caption{Consider the calculation of $t_1$ and $t_2$ such that the polyline
$p_0(t_0)p_1(t_1)p_2(t_2)p_3(t_3)$ is fully contained in ${\bf g}$.}
\label{fig:Order6}
\end{figure}

{\bf OP3} can be formalized by a system of three PDEs,
involving parameters
$t_i \in \R$ for critical edges $e_i$ of the step set.
The result ensures that $p_i(t_i)$ is the optimum 
point on $e_i$. Considering the situation illustrated in
Figure~\ref{fig:Order6}, this is equivalent to the
problem of finding the roots of $p(x) = 
 84x^6 - 228x^5 + 361x^4 + 20x^3 + 210x^2 + 200^x + 25$.
In fact, this problem
is not solvable by radicals over the field of rationals;
see \cite{LI2006}. (The proof uses a theorem by C. Bajaj \cite{BAJ1988} and the
factorization algorithm by E.\ R.\ Berlekamp \cite{BER1970}.)

This example allows two corollaries. Obviously and well-known is that
there is no exact arithmetic algorithm 
for calculating the roots of polynomials (known since
E. Galois; B.L. van der WaerdenÕs famous example is $p(x) = x^5 - x - 1$).
And secondly, there is also no exact arithmetic algorithm
for calculating 3D ESPs. C.\ Bajaj \cite{BAJ1985} showed this
based on a polynomial of order 20 for the general 3D ESP problem.
As a new result, here we have an oder 6 polynomial, and the
restricted ESP problem for simple cube-curves!

Note that this is not just a ``rounding number problem''  but a
fundamental non-existence of exact algorithms,
no matter what kind of time-complexity is allowed.

There is a uniquely defined shortest path, which 
passes through subsequent line segments $e_1, e_2,\ldots, e_k$ in 3D space in 
this order; see, for example, \cite{CHO1994}.
Obviously, vertices of a shortest path can be 
at real division points, and even at those which 
cannot be represented by radicals over the field 
of rationals.

\section{Approximate Algorithms}

An algorithm is an  $(1+\varepsilon)$-{\it approximation algorithm} for 
a minimization problem $P$ iff, for each input instance 
$I$ of $P$, the algorithm delivers a solution that is at 
most  $(1+\varepsilon)$  times the optimum solution \cite{HOC1997}.

The general 3D ESP problem can be solved
in  ${\cal O}\left(n^4\left[b + \log(n/\varepsilon)\right]^2/\varepsilon^2\right)$                                     time by an $(1+\varepsilon)$-approximation algorithm; see    
C. H. Papadimitriou \cite{PAP1985}.

An algorithm is  {\it $\kappa$-linear}  iff its time complexity is 
in  $\kappa(\varepsilon) \cdot {\cal O}(n)$, and function  $\kappa$   does 
not depend on  the problem size $n$, for  $\varepsilon  > 0$.
We use  $\kappa(\varepsilon)=({\cal L}_0 - {\cal L})/\varepsilon$,  where ${\cal L}$ is 
the true length of the ESP, and ${\cal L}_0$ the initial length.

A cube-curve is {\it first-class} iff each critical edge
contains one ESP vertex. The original RBA is correct
and   $\kappa$-linear for first-class cube-curves \cite{LI2006}.

\begin{figure}[!t]
\begin{center}
\vspace{-3mm}
\includegraphics[width=55mm]{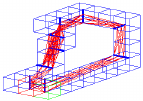}
\vspace{-3mm}
\end{center}
\caption{Weighted undirected graph for $m=3$.}
\label{fig:m3}
\end{figure}

We analyzed the following approximate graph-theoretical algorithm:
Subdivide each critical edge by $m$ uniformly-spaced
vertices; connect each vertex with those vertices such that 
the resulting edge is contained in the tube ${\bf g}$. This defines
a weighted undirected graph (see Figure~\ref{fig:m3}). Calculate a shortest-length cycle,
and use this as a (first-class !) input for the original RBA.
%
%

The time-complexity of the graph-theoretic algorithm (in our specification)
equals ${\cal O}\left(m^4n^4 + \kappa(\varepsilon)\cdot n\right)$.
It  applies DijkstraÕs algorithm repeatedly; possibly its time-complexity
can be reduced, but certainly not to be $\kappa$-linear. 

However, this (slow) algorithm allowed
for the first time to evaluate results obtained by the
original RBA.

Assume a simple cube-curve $g$ and a triple of 
consecutive critical edges $e_1$, $e_2$, and $e_3$ such that 
$e_i$ is orthogonal to $e_j$, for $i, j = 1, 2, 3$ and $i\ne j$. 
If $e_1$ and $e_3$ are also coplanar, then we say that $e_1$, $e_2$, 
and $e_3$ form an {\it end angle}, and a {\it middle angle} otherwise.

The following approximate numerical algorithm
requires an input  which is first-class
and has at least one end angle; the cube-curve is split 
at end angles into one or several arcs.
For each arc, one vertex on each critical edge
can be calculated using the systems of PDEs
briefly mentioned already above;
variable $t_i$ determines
the position of vertex $p_i$ on edge $e_i$.
This algorithm is provably correct and $\kappa$-linear
for the assumed inputs.

\begin{figure}[!t]
\begin{center}
\includegraphics[width=85mm]{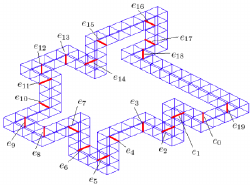}
\end{center}
\caption{A simple cube-curve where the ESP does not have
any grid-point vertex (and which has no end angle).}
\label{fig:noEA}
\end{figure}

An open problem in \cite{KLE2004} (page 406)
was stated as follows: Is there a simple cube-curve
such that none of the vertices of its ESP is a grid vertex?
The answer is ``yes'' \cite{LI2006}, and any of those curves
does not have any end angle; see Figure~\ref{fig:noEA}.\footnote
{
   Here are two new open problems:
   What is the smallest (say, in number of cubes
   or in number of critical edges - both is equivalent)
   simple cube curve which does not have 
   any end angle?
   What is the smallest (say, in number of cubes
   or in number of critical edges - both is equivalent)
   simple cube curve which does not have 
   any of its MLP vertices at a grid point location?
   We assume that the second problem is more difficult
   to solve.
}
Thus, the provably correct approximate
numerical algorithm cannot be used in general.
%

\begin{figure}[!b]
\begin{center}
\includegraphics[width=85mm]{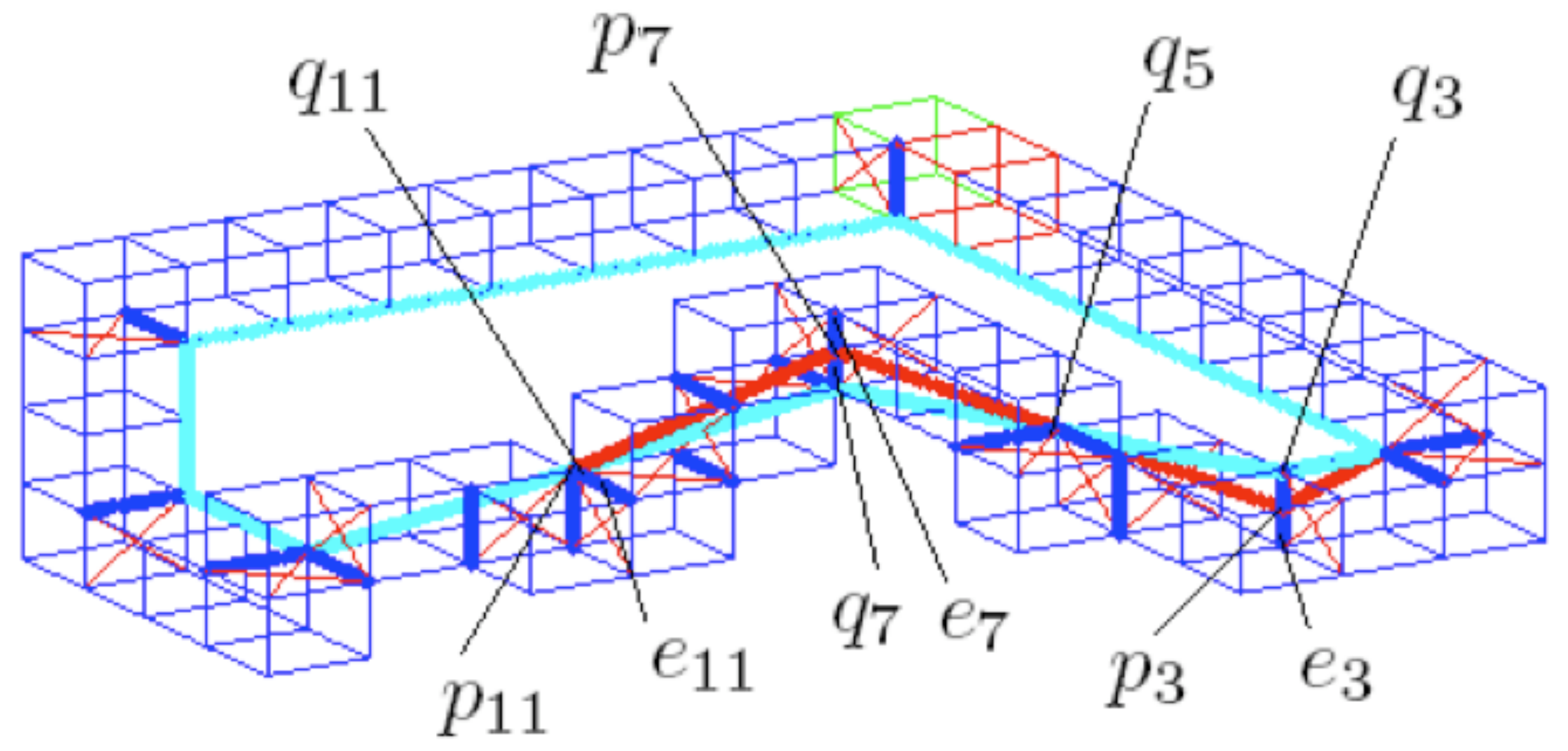}
\end{center}
\caption{A situation where the original Option 2 fails.}
\label{fig:NewOp2}
\end{figure}

This lead us back to the initial two questions about the original RBA: 
is it correct? (We can use either the approximate graph-
theoretical or the numerical algorithm for evaluation.)
What is its time-complexity in general?
Indeed, corrections were in place:

{\bf OP2}: if intersecting with the triangle $p_{i-1}p_ip_{i+1}$ 
         and using the convex arc only, we may miss 
         edges of the step set (see Figure~\ref{fig:NewOp2} for 
         such a situation) - more tests are
         needed, and this option was totally reformulated
         (for details, see \cite{LI2006} - the specifications require
         some technical preparations which cannot be given in
         this short paper).

{\bf OP3}: the vertex $p_{new}$, found by optimization, may
         specify edges $p_{i-1}p_{new}$ and $p_{new}p_{i+1}$ such that 
         one or both of them are not fully contained in 
         the tube of the curve; an additional test is
         needed (a simple correction).

Thus, as an end to this story right now, those
corrections define a provably correct  (for any simple cube-curve)
and $\kappa$-linear {\it edge-based RBA} \cite{LI2006}.

Instead of moving points along critical edges, we can
also move points within {\it critical faces} (which contain 
one critical edge). Of course, the vertices will finally move
onto or towards critical edges. This conceptually simpler
(in its {\bf OP2}) {\it face-based RBA} is also provably
correct, but showing a slower
convergence (within the limits of being $\kappa$-linear)
towards the EPS.

\begin{figure}[!t]
\begin{center}
\vspace{-3mm}
\includegraphics[width=65mm]{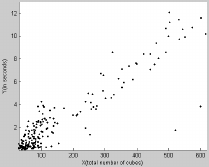}
\vspace{-3mm}
\end{center}
\caption{Edge-based RBA Implemented in Java, run under Matlab 7.0.4, Pentium 4,
using $\varepsilon = 10^{-10}$.}
\label{fig:time}
\end{figure}

See Figure~\ref{fig:time} for some statistics about measured run time.
Half of a simple cube-curves was generated randomly, and the second half then
generated using three straight arcs for closing the curve. The number of
cubes in generated curves was between 10 and 630. The break-off criterion
was defined by $\varepsilon = 10^{-10}$.

\begin{figure}[!b]
\begin{center}
\vspace{-4mm}
\includegraphics[width=65mm]{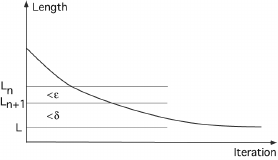}
\vspace{-5mm}
\end{center}
\caption{Let  $\varepsilon$  be the maximum accuracy of the
program, that means the smallest number for discriminating between
${\cal L}_n$ and ${\cal L}_{n+1}$. Still, the difference to the true value
${\cal L}$ might be $\delta > \varepsilon$.
The algorithm allows to obtain arbitrary accuracy
(with respect to ${\cal L}$) when continuing iterations, but this would require
to reduce $\varepsilon$.}
\label{fig:Cauchy}
\end{figure}

Figure~\ref{fig:Cauchy}  illustrates the meaning of the break-off criterion.
The lengths ${\cal L}_n$, for loops $n=1,2,3\ldots$ define a Cauchy sequence
which converges towards the true length ${\cal L}$. An in-depth study of this sequence
may reveal whether we can assume $\delta < \varepsilon$ in general, or not.

\section{ Conclusions }

This paper reported about the process of solving one particular
ESP problem. The developed methodology [i.e., define ``critical'' subsets,
specify the step set such that each critical subset in this set contains exactly 
one (possibly redundant, such as colinear) vertex, apply {\bf OP3}]
can be applied to ESP problems as considered (e.g.) in
\cite{MIT2004}. A few RBA applications have been illustrated in
\cite{LI2006,LI2007}, and further applications will be published soon
by the authors.


\begin{thebibliography}{99}

\bibitem{BAJ1985}
C.\ Bajaj.
The algebraic complexity of shortest paths in polyhedral spaces.
In Proc. \emph{Allerton Conf. Commun. Control Comput.},
pages 510--517, 1985.

\bibitem{BAJ1988}
C.\ Bajaj.
The algebraic degree of geometric optimization problems.
{\it Discrete Computational Geometry},
{\bf 3}:177--191, 1988.

\bibitem{BER1970}
E.\ R.\ Berlekamp.
Factoring polynomials over large finite fields.
{\it Math. Comp.},  {\bf 24}:713--735, 1970.

\bibitem{BUE2002}
T.\ B{\"u}low and R.\ Klette.
Digital curves in 3D space and a linear-time length estimation
algorithm.
{\it IEEE Trans.\ Pattern Analysis Machine Intelligence},
{\bf 24}:962--970, 2002.

\bibitem{CAN1987}
J. \ Canny and J.\ H. \ Reif.
New lower bound techniques for robot motion planning problems.
In Proc. \emph{IEEE Conf. Foundations Computer Science},
pages 49--60, 1987.

\bibitem{CHA1991}
B.\ Chazelle.
Triangulating a simple polygon in linear time.
{\it Discrete Computational Geometry}, 
{\bf 6}:485--524, 1991.

\bibitem{CHO1994}
J. \ Choi, J. \ Sellen, and C.-K. \ Yap.
Approximate Euclidean shortest path in 3-space.
In Proc. \emph{ACM Conf. Computational Geometry},
ACM Press, pages 41--48, 1994.

\bibitem{HOC1997}
D.\ S.\ Hochbaum (editor).
{\it Approximation Algorithms for NP-Hard Problems}.
PWS Pub. Co.,Boston, 1997.

\bibitem{KLE1999}
R.\ Klette, V.\ V.\ Kovalevsky, and B.\ Yip.
Length estimation of digital curves.
In Proc. {\it Vision Geometry}, SPIE 3811, pages 117--129, 1999.

\bibitem{KLE2000}
R.\ Klette and T.\ B{\"u}low.
Critical edges in simple cube-curves.
In Proc. \emph{Discrete Geometry Computational Imaging}, 
LNCS 1953, pages 467--478, 2000.

\bibitem{KLE2004}
R.\ Klette and A.\ Rosenfeld.
{\it Digital Geometry}.
Morgan Kaufmann, San Francisco, 2004.

\bibitem{LI2006}
F.\ Li and R.\ Klette.
Exact and approximate algorithms for the calculation of shortest paths. 
Report 2141 on\\
{\it www.ima.umn.edu/preprints/oct2006}

\bibitem{LI2007} 
F.\ Li and R.\ Klette.
Rubberband algorithms for solving various 2D or 3D shortest path problems.
In Proc. {\it Computing: Theory Applications}, plenary talk, pages 9--19, 2007.

\bibitem{MIT2004}
J.\ S.\ B. \ Mitchell and M. \ Sharir.
New results on shortest paths in three dimensions.
In Proc. \emph{SCG}, pages 124--133, 2004.

\bibitem{PAP1985}
C.\ H.\ Papadimitriou. 
An algorithm for shortest path motion in three dimensions.
{\it Inform. Process. Lett.}, {\bf 20}:259--263, 1985.

\bibitem{TAL2004}
M.\ Talbot.
A dynamical programming solution for shortest
path itineraries in robotics.
{\it Electr. J. Undergrad. Math.},
{\bf 9}:21--35, 2004.

\end{thebibliography}
\end{document}